\documentclass[runningheads]{llncs}
\usepackage[T1]{fontenc}

\usepackage{graphicx}
\usepackage{caption}
\usepackage{xurl}
\usepackage{import}
\usepackage[misc,geometry]{ifsym}
\usepackage{hyperref}
\begin{document}
\title{Effects of the Cyber Resilience Act (CRA) on Industrial Equipment Manufacturing Companies}
\titlerunning{Effects of CRA on Industrial Equipment Manufacturing Companies}

\author{Roosa Risto \Letter \orcidID{0009-0005-3138-913X} \and \\
    Mohit Sethi\orcidID{0000-0002-9730-1955} \and \\
    Mika Katara\orcidID{0009-0003-5512-9442}}
%
\authorrunning{R. Roosa et al.}
%
\institute{KONE Corporation\\\email{firstname.lastname@kone.com}}

\maketitle              

\begin{center}
\textbf{Accepted Manuscript} \\
This version of the article has been accepted for publication in ARES 2025 workshop. The final version is available online at SpringerLink.
\end{center}

\begin{abstract}
    The Cyber Resilience Act (CRA) is a new European Union (EU) regulation aimed at enhancing the security of digital products and services by requiring them to meet stringent cybersecurity requirements. To understand the practical implications of CRA for industrial equipment manufacturing companies, a survey was conducted to identify key challenges. The results revealed significant hurdles, including the implementation of secure development lifecycle practices, managing vulnerability notifications within strict timelines, and addressing gaps in cybersecurity expertise. Based on these findings, the paper offers targeted recommendations in key focus areas such as vulnerability management and tooling improvements to support industrial equipment manufacturers in preparing for CRA compliance.\footnote{This is the accepted manuscript version. The final published version is available at \url{https://doi.org/10.1007/978-3-032-00630-1_12}}
\end{abstract}

\section{Introduction}
Motivated by the growing number of cyberattacks targeting connected devices and the lack of consistent baseline cybersecurity practices among manufacturers, the European Union (EU) Cyber Resilience Act (CRA)~\cite{CRA2024} sets common security requirements for all products with digital elements within the EU. This broad scope intentionally covers all software and hardware products, as well as their remote data processing solutions, with exceptions for medical devices, aeronautical equipment, etc. which are already subject to similar existing EU regulations. The primary goal of CRA is to enhance cybersecurity across the European Union by ensuring that products with digital elements are designed, developed, and maintained with robust security measures. While the act imposes requirements on many stakeholders, a significant portion of these requirements are directed towards manufacturers. The requirements for manufacturers can broadly be classified into software development lifecycle (SDL) process requirements, technical requirements, vulnerability management and notification requirements, and preparation of technical and user documentation.

The \textbf{SDL and technical} requirements outlined in Annex I Part I of the act require manufacturers to perform a risk assessment and subsequently ensure an appropriate level of cybersecurity throughout the design, development, and production process based on the risk assessment. SDL practices must ensure that the products are free of known exploitable vulnerabilities, configured securely by default, and support timely security updates, including automatic updates with opt-out options. Additional technical measures include protecting data confidentiality, integrity, and availability through mechanisms like encryption and access controls, minimizing attack surfaces, implementing resilience against denial-of-service attacks, and data minimization. Furthermore, technical measures must also ensure logging of security-relevant activity, possibility of secure data removal, and mitigation of negative impact on other devices or networks.

Annex I Part II of the CRA sets forth \textbf{vulnerability management} requirements, obligating manufacturers to identify and document components and vulnerabilities in their products, including the creation of a software bill of materials (SBOM) in a standard, machine-readable format. Manufacturers are also required to address and remediate vulnerabilities promptly, providing security updates as necessary. When releasing a security update, the manufacturers must publicly disclose details about the fixed vulnerabilities, such as impacts, severity, and remediation steps. Lastly, manufacturers must perform regular testing to identify new vulnerabilities.

CRA specifies \textbf{vulnerability notification} practices in articles 14 and 15, mandating manufacturers to inform the European Union Agency for Cybersecurity (ENISA) and national computer security incident response teams (CSIRTs) within 24 hours of identifying actively exploited vulnerabilities or severe cyber incidents. The initial notification is followed by a detailed report within 72 hours, and a final report within 14 days for actively exploited vulnerabilities and within one month for other incidents. Manufacturers are also encouraged to voluntarily report near misses or incidents impacting the security of their products.

\textbf{User documentation} requirements listed in Annex II of CRA expect manufacturers to provide documentation containing, among other things, contact details for reporting vulnerabilities, the name and type information for uniquely identifying the product, information on the intended purpose of the product, detailed guidance on secure commissioning, update installation, and decommissioning, the type and duration of technical security support, and optionally, access to the product SBOM.

Requirements in Annex VII of CRA instruct that \textbf{technical documentation} must include a general description of the product, its intended purpose, relevant software versions, and, for hardware products, visual representations like photographs or diagrams of external and internal interfaces. Additionally, the technical documentation should detail the cybersecurity risk assessments performed, the SDL practices followed, the system architecture designed and developed, a list of applicable standards with which the product is compliant, as well as test reports verifying conformity with the technical requirements in those standards, and potentially the product SBOM.

\begin{figure}[!h]
    \centering
    \includegraphics[width=\linewidth]{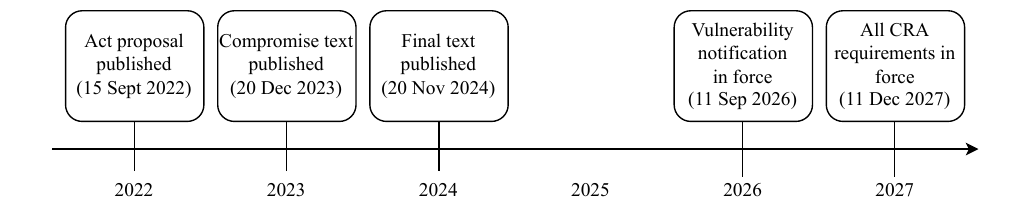}
    \caption{CRA timeline}
    \label{fig:timeline-vector}
\end{figure}

Figure~\ref{fig:timeline-vector} shows the timeline for when different CRA requirements come into force. The final CRA text was published on 20 November 2024, implying that the transition period for vulnerability notification responsibilities will end on 11 September 2026, and all CRA requirements will become mandatory from 11 December 2027.

Non-compliance with the CRA can have serious consequences for manufacturers, including restricted market access, mandatory product recalls, or financial penalties. Given the extensive regulatory obligations and the relatively short timeline for full compliance, we conducted a survey to identify the specific challenges faced by industrial equipment manufacturing companies. These companies typically have well-established engineering practices, large development organizations, and structured product lifecycle processes already in place. However, their products often include deeply embedded digital elements integrated into complex, safety-critical, bespoke system installations, such as cranes and mining equipment. In such contexts, supporting requirements like automatic security updates and configuration changes are non-trivial. Additionally, organizational inertia and the need to coordinate changes across multiple product lines, regulatory environments, and customer expectations can make the adoption of CRA particularly challenging. Our goal was to identify concrete pain points and offer targeted recommendations to aid in their preparation for compliance.

The rest of this paper is organized as follows. Section~\ref{sec:related} reviews prior work on the challenges associated with CRA implementation. Section~\ref{sec:iec62443} examines the IEC 62443 series of standards and discusses how they align with CRA requirements. The study design, including participant background and familiarity, is described in Section~\ref{sec:study}. Section~\ref{sec:challenges} presents the key challenges identified by the surveyed industrial equipment manufacturers. In Section~\ref{sec:recommendations}, we provide targeted recommendations to address the challenges identified. Finally, Section~\ref{sec:conclusion} concludes the paper and summarizes the main findings.

\section{Previous Work on CRA Challenges}
\label{sec:related}
Szedlak et al.~\cite{Szedlak2024} examine the awareness and preparedness of Small and Medium-sized Enterprises (SMEs) for CRA. Their findings reveal significant disparities in CRA awareness and readiness, with only 12.3\% of SMEs being aware of the CRA compared to 83.5\% of very large enterprises. 

Thiel~\cite{thiel2024challenges} in his article discusses challenges in CRA compliance, particularly ensuring third-party components being integrated into products also meet CRA standards. He also highlights the issue of lacking in-house cybersecurity expertise, which may necessitate the need for external support. 

The European Cyber Security Organisation (ECSO) conducted an online survey to identify potential challenges linked to the implementation of CRA~\cite{ECSO2024}. The survey participants included a subset of ECSO members, primarily security service providers and manufacturers of digital devices, along with other public and private organizations. The survey revealed several key challenges, including the lack of clarity with product categories, the proposed timeline for implementation, and the need for clear guidelines on conducting risk and conformity assessments. Respondents emphasized the necessity for harmonized standards and methodologies at the EU level, as well as the importance of guidelines, templates, and training to facilitate CRA implementation.

A joint letter by experts published by the Center for Cybersecurity Policy and Law~\cite{experts2023joint} raised concerns about the CRA's vulnerability disclosure requirements, including risks of misuse by governments, potential compromise of the vulnerability database by malicious actors, and discouraging effect on collaboration among manufacturers and researchers. The letter emphasizes the need for a risk-based approach, considering factors such as the severity of the vulnerability, the availability of mitigations, the potential impact on users, and the likelihood of broader exploitation.

A blog post by the Open Source Initiative~\cite{open_source_list} captures feedback on the draft text of CRA from various organizations in the open source community, highlighting concerns such as the unclear scope of exemptions for open source software, the risk of imposing disproportionate regulatory burdens on non-commercial contributors, and potential disruption to collaborative development models. 

Schoo~\cite{Schoo2024} in his paper examines some of the challenges from the Cyber Resilience Act (CRA). Among other things, he highlights the need for establishing uniform cybersecurity standards across the European Union and the necessity for a significant increase in the number of conformity assessment bodies to effectively manage and enforce these standards.

While previous work highlighted several challenges associated with CRA, our survey focuses on investigating the specific challenges that CRA poses for industrial equipment manufacturing companies. By identifying these challenges, we aim to provide an understanding of how CRA impacts industrial equipment manufacturers and offer insights into effective strategies in key focus areas for seamless CRA compliance.

\section{CRA and IEC 62443}
\label{sec:iec62443}
CRA~\cite{CRA2024} in Annex VIII offers four conformity assessment options, including internal control, where manufacturers self-declare conformity, and EU-type examination, where a notified body verifies conformity before issuing a certificate. Harmonized standards play a pivotal role in all assessment options. As noted in recital 79 of the act, harmonized standards translating CRA's requirements into detailed technical specifications, offer manufacturers clear and actionable compliance guidelines. Recital 81 further underscores that products with digital elements certified or declared in conformity with harmonized standards will benefit from a presumption of conformity, streamlining the compliance process. 

The IEC 62443 series of standards provides a comprehensive framework for securing industrial automation and control systems (IACS) against cybersecurity threats. This series of standards is likely to form the basis for any harmonized standard used for CRA compliance by industrial equipment manufacturing companies. This is also evident from the ECSO survey on CRA challenges~\cite{ECSO2024}, which noted IEC 62443 as one of the most quoted standards by survey participants. While CENELEC, the European electrotechnical standards body, and specialized industrial equipment manufacturers may develop other harmonized standards, they will still reuse the IEC 62443 series as the building block. For example, the ISO 8102-20~\cite{ISO8102-20} standard for elevators is based on IEC 62443.

The IEC 62443 standard series is structured into multiple parts, addressing different aspects of security, from general concepts and policies to technical requirements and processes. Among these, the parts relevant for CRA requirements include IEC 62443-4-1~\cite{IEC62443-4-1}, which outlines secure product development lifecycle (SDL) requirements for vendors; IEC 62443-4-2~\cite{IEC62443-4-2}, which specifies technical security requirements for IACS components, such as controllers and software applications; and IEC 62443-3-2~\cite{IEC62443-3-2}, which focuses on security risk assessment and system design. The relevance of these standards for CRA compliance is also captured in the standard mapping report prepared by the European Union Agency for Cybersecurity (ENISA)~\cite{ENISA2024}.

In our survey, a major component of the questions focused on the IEC 62443-4-1 secure development lifecycle (SDL) process requirements relevant to CRA compliance. Organizations that have already adopted IEC 62443 practices are generally well positioned to meet many of the CRA's expectations, especially in areas related to secure development, risk-based design, and technical protections for the confidentiality and integrity of transmitted and stored data. Many of the participating companies were indeed applying IEC 62443 development practices internally. However, CRA introduces additional requirements not fully covered by the current IEC 62443 series. These include fixed timelines for vulnerability notification, mandatory public disclosure under specific conditions, and more detailed user documentation expectations. Therefore, in addition to questions on SDL process challenges, our survey also explored participants’ views on meeting CRA’s vulnerability notification obligations and user documentation requirements, which extend beyond what is explicitly addressed in IEC 62443. To deliver CRA-compliant products by December 2027, manufacturers will need to develop and integrate these additional capabilities into their existing processes.

\section{Study Design}
\label{sec:study}
Between June and August 2024, we conducted a survey to gather insights into the compliance challenges faced by industrial equipment manufacturing companies. The survey utilized a detailed online questionnaire comprising 44 questions that covered various aspects of CRA and IEC 62443-4-1 compliance. The questionnaire included both multiple-choice and freeform questions, focusing on topics such as the role of EU standards within the organization, familiarity with IEC 62443 and CRA, existing secure software development lifecycle (SDL) practices, whether companies already follow IEC 62443-4-1, perceived difficulty in implementing specific practices, challenges in meeting CRA requirements such as vulnerability management and notification, expected role and budget changes, and anticipated compliance timelines.

Participants completed the questionnaire during one-hour video meetings, where they could provide additional feedback using the think-aloud protocol. This approach helped capture verbal comments that might otherwise have been missed, especially in response to the freeform qualitative questions. All meetings were hosted by the same author, and the questionnaire was shared with participants prior to the video call. Each session concluded with an open-ended discussion to gather broader feedback on challenges and potential solutions. Participants were free to refer to the IEC 62443 standard and the CRA text while answering the questions. If they asked for clarification about the meaning of specific requirements or practices, the facilitator provided contextual explanations and examples to ensure a shared understanding.

A total of 12 companies participated in the survey. This group included seven large companies (with over 500 employees) that manufacture industrial equipment such as cranes, elevators, and machinery for the shipping and mining industries. The remaining five companies were consultancies or partner organizations providing software and support services to the large industrial equipment manufacturers we surveyed. For each company, we ensured that the survey was answered by a senior technical cybersecurity manager who had a comprehensive understanding of both the organizational perspective and the technical challenges. Only one response per company was allowed, although multiple cybersecurity experts from the organization could join the calls. In fact, for two of the 12 companies, more than one expert participated in answering the questions. All participating companies had a sizable portion of their revenues in Europe, which motivated their interest in maintaining compliance in the region.

The survey responses were analyzed using a qualitative coding approach commonly used in exploratory user studies. One of the authors reviewed the meeting recordings and annotated relevant open-ended comments that were not captured in the survey form. These responses were then coded thematically to identify recurring topics and pain points. Both text-based and spoken responses were included in the frequency analysis to determine how often specific concerns or challenges were mentioned across participants. Observed patterns and consistent themes were synthesized to form the basis for the practical recommendations presented later in the paper.

During the survey, we used the CRA text from December 2023~\cite{cra_2023_version}. Participants were aware that the final text would be published in the first half of 2024, with a 21-month transition period for notification responsibilities ending in 2026, and a 36-month period for other responsibilities ending in 2027. Given that these expected timelines were similar to the final dates, it is unlikely that the survey responses were significantly affected.

\begin{figure}[htbp] 
    \centering 
\includegraphics[width=\linewidth]{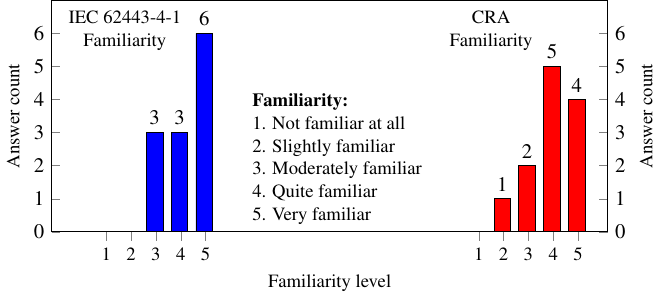}
  \caption{Participant familiarity with IEC 62443-4-1 and CRA.} 
    \label{fig:familiarity} 
\end{figure}

\textbf{Baseline Understanding} Before delving into questions about compliance challenges, we wanted to gauge the familiarity of the participants with the IEC 62443-4-1 standard and the relatively new CRA regulation. As shown in Figure~\ref{fig:familiarity}, most participants were familiar with both IEC 62443-4-1 and CRA, with 75\% (9 out of 12) companies being quite or very familiar. Given that the IEC 62443 series of standards has been around for several years, it is not surprising that participants were more familiar with it compared to CRA. While participants demonstrated familiarity with both the standard and the regulation, they were allowed and encouraged to consult the official text during the survey and if any participant had questions about specific requirements, we provided assistance by referring to the relevant sections in the text.

IEC 62443-4-1 includes different levels of maturity: Maturity Level 1 (Initial), Maturity Level 2 (Managed), Maturity Level 3 (Defined/Practiced), and Maturity Level 4 (Improving). All of the industrial equipment manufacturing companies interviewed had already established an internal SDL compliant with IEC 62443-4-1, and a few of them had reached Maturity Level 3, meaning they were actively practicing the SDL. Several manufacturers had even undergone third-party audits and obtained certifications. Naturally, the consulting and partner companies followed the SDL practices of the equipment manufacturing companies they worked with to ensure their customers remained compliant but were not developing their own SDL.

\textbf{Limitations} We recruited survey participants based on their involvement in a joint research project focused on automating compliance with the IEC 62443 series of standards. Additional participants were invited through recommendations from those already involved. All participating companies were predominantly based in Northern Europe, which may limit the generalizability of the findings to all industrial manufacturers subject to the CRA. Although some participants were international, with products sold outside the EU and not subject to CRA, the challenges faced by manufacturers outside the EU but selling products in the EU may differ. Another limitation is the small sample size of 12 companies, which makes it inappropriate to draw statistically significant conclusions. This limitation highlights the qualitative nature of our study, which is primarily descriptive and exploratory rather than statistically validated. Nonetheless, the findings provide valuable insights into the challenges faced by industrial equipment manufacturers. Since the focus was on identifying challenges specific to industrial equipment manufacturers, the recommendations may not apply to manufacturers of other products, such as consumer electronics.
\section{Challenges}
\label{sec:challenges}
\begin{figure}[htbp] 
    \centering 
\includegraphics[width=\linewidth]{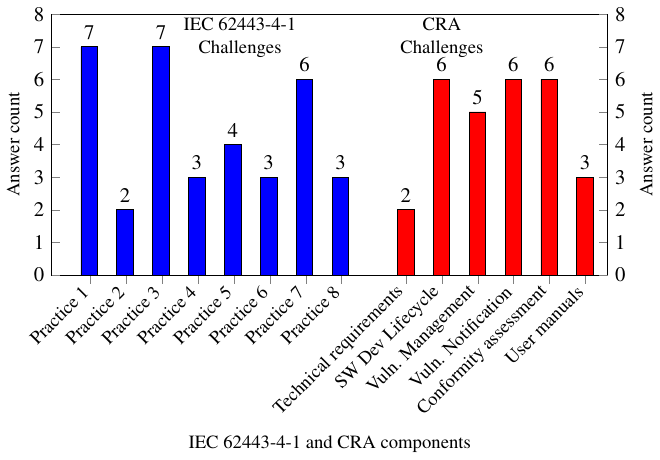}
  \caption{Challenges with IEC 62443-4-1 and CRA requirements.} 
    \label{fig:challenges} 
\end{figure}

IEC 62443-4-1 organizes the SDL requirements into eight distinct practices, covering various aspects of the development lifecycle, which are listed in table~\ref{tab:iec_practices_table}. The survey asked participants to select any practice that they found challenging to incorporate into their SDL. Participants were free to choose multiple practices when selecting those they found difficult. The survey responses for the practices that participants found challenging to implement are illustrated in figure~\ref{fig:challenges}.

\begin{table}[!ht]
    \centering
    \begin{tabular}{|l|l|}
    \hline
        Practice 1 & Security management \\ \hline
        Practice 2 & Specification of security requirements \\ \hline
        Practice 3 & Secure by design \\ \hline
        Practice 4 & Secure implementation \\ \hline
        Practice 5 & Security verification and validation testing \\ \hline
        Practice 6 & Management of security-related issues \\ \hline
        Practice 7 & Security update management \\ \hline
        Practice 8 & Security guidelines \\ \hline
    \end{tabular}
    \caption{IEC 62443-4-1 practices}
    \label{tab:iec_practices_table}
\end{table}

As shown in figure~\ref{fig:challenges}, survey participants identified security management, secure by design, and security update management practices as the most challenging. Participants who selected a practice as difficult to implement had the opportunity to elaborate on their specific challenges through a freeform text question and verbal explanations during the meeting. Several recurring themes emerged from their responses.

After selecting the security management practice as a challenge, the participants voiced several concerns. This practice requires establishing a comprehensive SDL as part of the product development process and enforcing it across the organization for both existing and new projects makes it inherently complex. Another challenge highlighted was determining the applicability of specific requirements in the standard. For instance, participants noted difficulty in deciding when requirements, such as providing hardening guidelines, apply to their unique circumstances as industrial equipment manufacturers since often end users are not responsible for installation, hardening, or maintenance of industrial equipment on their own. Another significant issue was identifying and allocating responsibilities, as organizational structures often do not align with the process structures outlined in the standard. This misalignment necessitates negotiating additional responsibilities for teams and securing management buy-in at various levels. Communication gaps between management, developers, and cybersecurity teams were also frequently mentioned. Upper management often lacks sufficient understanding of cybersecurity requirements, while cybersecurity specialists sometimes struggle to effectively convey challenges and issues to leadership, leaving participants uncertain about how to address these communication barriers. Additionally, the practice mandates ensuring adequate security expertise and it was seen as a challenge, with participants citing a perceived shortage of qualified cybersecurity professionals. A perhaps subtle challenge that might have been missed is also the difficulty in evaluating whether personnel assigned to roles have the appropriate cybersecurity skills. Interestingly, participants did not report difficulties with the more technical controls within the security management practice 1, such as SM-6 (file integrity) or SM-8 (controls for private keys), suggesting that technical controls are generally perceived as less challenging compared to organizational and procedural aspects.

Participants identified a lack of cybersecurity-educated or experienced experts as a key challenge with the secure by design practice 3. They noted the difficulty of finding seasoned professionals who possess a broad understanding of defense-in-depth practices, such as least privilege, audit logging, and secure boot, while also being capable of collaborating with multiple product teams. Additionally, these experts must be able to dedicate sufficient time to thoroughly document design reviews and assessments, further complicating the implementation of this practice.

Security update management practice evoked conversation on several challenges that survey participants face when aiming to comply. This is understandable, as many industrial equipment manufacturers rely on embedded software packages that are delivered as complete images, unlike desktop software libraries where individual patches can be built, delivered, and applied to address specific vulnerabilities. Furthermore, industrial equipment often has intermittent or no connectivity, making it difficult to meet timely update requirements. While the standard acknowledges that the availability and safety properties of industrial equipment take precedence over addressing theoretical vulnerabilities, participants still found compliance with the update management practice challenging. Finally, the end users of industrial equipment are often not qualified to access or apply updates. This creates a disconnect between the standard's requirements for security-update documentation and the practical realities of end-user capabilities.

\textbf{Challenges in CRA Implementation} 

When exploring challenges with CRA compliance, we categorized the act's requirements into technical requirements, secure development lifecycle (SDL) requirements, vulnerability management, vulnerability notification, conformity assessment, and user manuals. Technical documentation, including risk assessments and interface descriptions, was not queried as a separate category but was considered part of a mature SDL. Participants were allowed to select multiple answers to indicate the areas they found challenging. The responses for each category of requirements are summarized in Figure~\ref{fig:challenges}.

The CRA requirement text \textit{“Products with digital elements shall be designed, developed, and produced in such a way that they ensure an appropriate level of cybersecurity based on the risks”} implies the need for a mature secure software development lifecycle (SDL). Many survey participants found this challenging, echoing the difficulties observed with the IEC 62443-4-1 requirements discussed earlier. Establishing and enforcing an SDL across an organization is a substantial undertaking. 

Participants were also uncertain whether their vulnerability management processes and tools were sufficiently mature to meet CRA expectations. Several respondents highlighted challenges such as generating accurate and up-to-date SBOMs, running and integrating appropriate vulnerability scanning tools, and effectively triaging and prioritizing vulnerabilities. Even when vulnerabilities are identified, issuing timely patches can be difficult in practice, particularly in industrial settings where the update cycles are inherently slower due to end-user demands for high availability and minimal downtime. These operational realities were identified as barriers to fulfilling CRA’s requirements for timely remediation and clear vulnerability reporting.

Similarly, vulnerability notification requirements were also highlighted as a significant challenge, as many participants had little prior experience with such practices. The CRA deadline of notifying within 24 hours of an actively exploited vulnerability was perceived as stringent. Some participants even expressed doubts about their ability to detect actively exploited vulnerabilities promptly. 

Conformity assessment was another major area of concern. Participants were unsure whether their products required external assessment by a notified body or if internal self-declaration would suffice based on their product type. Even when self-declaration was deemed acceptable, participants questioned whether their quality teams, which currently oversee self-declaration and CE marking, were adequately equipped to evaluate compliance with the cybersecurity requirements of the CRA. In addition, it was not clear to participants whether sector-specific standards based on IEC 62443 would be sufficient for demonstrating conformity under the CRA. For example, the maritime sector has adopted the International Association of Classification Societies (IACS) Unified Requirements UR E26~\cite{IACS-UR-E26} and UR E27~\cite{IACS-UR-E27}, which apply to new ships contracted from 1 January 2024 and draw on IEC 62443. However, it was unclear to participants whether compliance with or certification against these standards will fully meet all CRA requirements or be sufficient for self-declaration or assessment by a notified body.

\begin{figure}[htbp]
    \centering
   \includegraphics[width=\linewidth]{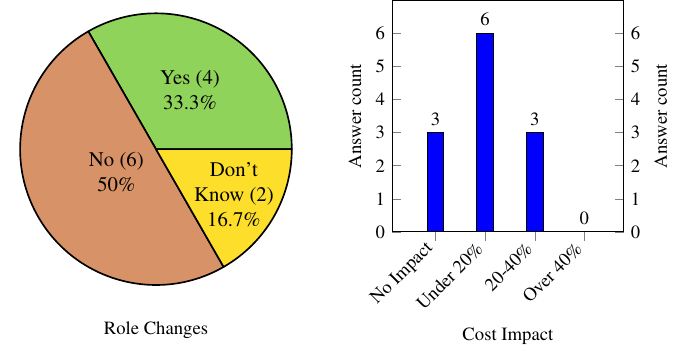}
    \caption{Anticipated Role Changes and Cost Impact of CRA.}
    \label{fig:role_changes_cost_impact}
\end{figure}

\textbf{Cost Impact and Organizational Changes} When asked about the estimated cost impacts of CRA compliance, as shown in Figure~\ref{fig:role_changes_cost_impact}, 50\% of participants expect costs to remain under 20\%, 25\% anticipate costs to increase by 20--40\%, and, surprisingly, 25\% foresee no cost impact at all. The question asked participants to estimate how much additional cost CRA compliance would introduce relative to their current product development budgets, including expenses related to personnel, tooling, and process adjustments. Similarly, when participants were asked about potential changes in employee roles, also shown in figure~\ref{fig:role_changes_cost_impact}, the majority (50\%) expect roles to remain unchanged after the CRA comes into force, while 33.3\% anticipate role changes, and 16.7\% remain uncertain. These findings suggest that CRA may not significantly impact employee roles or organizational costs. This could indicate that manufacturers are either reasonably prepared or have resigned themselves to achieving compliance using their existing resources without expecting additional budget or support for structural changes.

\textbf{Transition period} The questionnaire included a freeform question about whether the CRA transition period of 21 months for vulnerability notification requirements and 36 months for all other requirements is reasonable. The general consensus was that the transition period is reasonable, though some participants expressed concerns about uncertainty surrounding conformity assessment requirements. Additionally, some acknowledged that a longer transition period might not necessarily make compliance easier, as it could simply encourage companies to delay action until the last minute. One participant noted that while the transition period might seem reasonable on paper, it could prove insufficient without strong leadership commitment to the significant organizational changes required.

\textbf{Use of open-source software} To investigate any specific concerns related to open-source software, we included the following question in the survey: \textit{“What impacts do you see on the use of open-source software in your products that need to comply with the CRA?”} Participants acknowledged that complying with the CRA would make the use of open-source software more complex. They emphasized the need for accurate SBOM tracking tools to detect and document all open-source components comprehensively. Additionally, participants highlighted the need for more rigorous vetting processes to ensure components are actively maintained and the necessity of establishing reliable channels for receiving vulnerability information.

\textbf{Role of CRA in improving security posture} When asked about their overall impressions of how significantly the CRA is expected to enhance the cybersecurity of their products, participants provided varied responses. While the majority leaned toward the belief that the act would improve the security posture of their products, a few participants expressed skepticism, with some stating they did not anticipate any tangible improvements. This suggests that while many manufacturers recognize the potential benefits of the CRA, not all perceive it as a guaranteed driver of enhanced cybersecurity. Some may regard the act as more of a regulatory burden than a practical means to improve product security.

\section{Tackling CRA Challenges}
\label{sec:recommendations}
As is evident from the survey results, industrial equipment manufacturers anticipate several challenges in preparing for CRA compliance. This section focuses on the main areas where participants discussed practical difficulties and explores recommendations that emerged during the survey conversations to help address these challenges.

\textbf{Cybersecurity Experts} Several survey participants identified the shortage of qualified cybersecurity professionals as a critical challenge, particularly in the context of implementing IEC 62443-4-1 practices related to security management and secure-by-design principles. Article 10 of the CRA~\cite{CRA2024} acknowledges this issue but stops short of mandating specific actions. Instead, it encourages EU member states, ENISA, and the European Cybersecurity Competence Centre to develop \textit{“organizational and technological tools to ensure sufficient availability of skilled professionals.”} Addressing this shortage requires broad, long-term planning. However, in the interim, industrial equipment manufacturers must prioritize internal training, upskilling, and adopting a shift-left strategy. This approach integrates security responsibilities across all teams, rather than confining them to a dedicated cybersecurity team.

\textbf{Communication} Survey participants highlighted several communication challenges at various levels, including the need for clear and effective communication between the cybersecurity team and leadership to convey the CRA's extensive requirements, the number of affected products, required investment, management commitment, and the short transition timeline. Additionally, challenges arise between the cybersecurity team, which is well-versed in the SDL and technical requirements, and the quality team, which oversees self-declaration and CE labeling. The quality team will need to collaborate with cybersecurity to verify CRA compliance and ensure technical documentation is complete before signing a self-declaration. Lastly, participants raised concerns about communication with external authorities, such as ENISA and national CSIRTs, when making vulnerability notifications.

To address communication challenges, it is essential to establish regular cross-departmental meetings involving the cybersecurity team, company leadership, and the quality team. These meetings should ensure alignment on CRA requirements, investment needs, and clearly define the roles and responsibilities of each organizational unit in achieving CRA compliance. Additionally, setting clear next steps and timelines will help streamline the process and foster effective collaboration. Training sessions and workshops tailored for the quality team can further build a shared understanding of CRA requirements. Establishing dedicated communication channels and implementing feedback mechanisms will also help mitigate anticipated communication challenges. Adopting a shift-left strategy will integrate security considerations early in the development process, ensuring that security becomes a shared responsibility across all teams. To address communication challenges related to vulnerability notifications, ENISA and national CSIRTs should prioritize developing a unified Coordinated Vulnerability Disclosure (CVD) ecosystem with clear communication practices and automated tools for streamlined disclosures.

\textbf{Tooling} To encourage discussion, not only on the challenges but also on potential techniques to overcome them, our survey included an optional freeform question:~\textit{What kind of tooling would help your organization in fulfilling the CRA requirements and realizing its full benefits?} Participants had many suggestions, which we synthesized into four tooling categories, along with specific recommendations that refine their preliminary thoughts.

Several participants expressed a desire for advanced, cost-efficient \textit{DevSecOps} tools that support the entire development lifecycle, from threat modeling to requirements tracking, source code scanning, and penetration testing. While many tools are already available for such tasks, organizations should first study their existing development pipelines and understand how new projects and features are introduced and specified. It is crucial to ensure that the chosen DevSecOps tools can integrate seamlessly with these development pipelines. If the development pipelines do not align with modern DevSecOps principles, companies should reconsider and modify them accordingly.

The importance of \textit{vulnerability management} tools was a recurring theme among participants. Many recognized the value of standard SBOM formats like CycloneDX~\cite{CycloneDx} and tools for tracking vulnerabilities based on SBOM data. While vulnerability management tools are necessary and readily available, companies should also prepare for centrally aggregating vulnerabilities from multiple sources and constructing a hierarchy of vulnerabilities that accurately reflects the composition of their product components. Adopting a vulnerability risk scoring methodology grounded in Common Vulnerability Scoring System (CVSS) is essential too. To reduce noise from false positives or low-impact vulnerabilities, organizations should consider advanced risk-rating methods, including AI-driven analysis if necessary.

\textit{Governance Risk and Compliance (GRC)} tools were also mentioned by a couple of participants. While GRC tools can be helpful, the primary goal for industrial equipment manufacturing companies should be setting a company-wide Cybersecurity Management System (CSMS) aligned with IEC 62443 if it hasn't already been done. Once the CSMS is developed, it needs to be put into practice, and GRC tools can help operationalize and monitor its implementation. However, ensuring that all the policies in the CSMS are understood by employees and effectively applied across the organization is a significant challenge that cannot be solved by GRC tools alone. 

Participants also emphasized the necessity of \textit{incident response and workflow automation} tools to comply with CRA notification responsibilities. For this, ENISA and national CSIRTs should develop a unified CVD policy that supports the use of protocols such as Structured Threat Information eXpression (STIX)~\cite{STIX} to automate the exchange of threat and vulnerability information and standard vulnerability reporting formats like the Common Vulnerability Reporting Framework (CVRF)~\cite{CVRF}. Concurrently, organizations should prepare templates for initial, detailed, and final reports to streamline the notification process if and when an actively exploited vulnerability is discovered.

As a general recommendation for selecting tools to ensure CRA compliance, organizations should thoroughly evaluate the application programming interface (API) and integration support of each tool before adoption. This evaluation is crucial for building an end-to-end automated pipeline that spans from threat modeling to continuous penetration testing. Additionally, selecting tools with flexible licensing models that can scale with project needs is essential. Prior to making a commitment, organizations should pilot and verify that tools conform to organizational workflows. Lastly, engaging with peers at industry forums to share experiences and adopt best practices based on insights of others can enhance tool selection and implementation.

\textbf{Vulnerability Management and Notification Practices} In addition to tooling, manufacturers should develop and document internal vulnerability management processes that clearly define roles, responsibilities, and workflows from detection to resolution. These should cover vulnerability identification using scanning tools, risk triage to assess severity and impact, and escalation procedures for critical or high-priority findings. The process must also include handling false positives, assigning remediation tasks, and coordinating actions across relevant teams such as product owners, security engineers, and operations. It should define criteria for when public vulnerability notification is required under CRA, and assign roles to ensure timely and accurate reporting to authorities such as ENISA or national CSIRTs. The process should also account for compensating controls when immediate fixes are not feasible, and include steps for planning, approving, and deploying fixes through the organization's existing change management procedures. It must also support verification of fixes, handle residual risk through formal exception workflows, and be overseen by a central governance structure that regularly reviews critical findings and notification decisions. Where available, manufacturers are encouraged to base their process on harmonized standards that are intended to demonstrate CRA compliance within their sector. In the absence of such a standard, ISO/IEC 29147~\cite{ENISOIEC29147} and ISO/IEC 30111~\cite{ENISOIEC30111} can serve as useful inputs for designing an effective internal process.

\section{Conclusion}
\label{sec:conclusion}
We conducted a survey of industrial equipment manufacturing companies to explore the challenges they anticipate as they prepare for CRA compliance. The survey revealed several key concerns, including the tight timelines for vulnerability notifications, the requirement to support timely automatic security updates with opt-out options, the need for updated user manuals, the complexities associated with using open-source software, and uncertainty around conformity assessment procedures and the applicability of sector-specific standards. Through freeform discussions during the survey, participants shared ideas for addressing these challenges, which we have synthesized into practical recommendations across several focus areas. These include improving internal vulnerability management processes and notification readiness, enhancing communication between cybersecurity, quality, and leadership teams, addressing gaps in cybersecurity expertise through training and shift-left strategies, integrating supportive tooling into development workflows, and ensuring that conformity assessment processes are well understood and documented. While organizations already following IEC 62443 are better prepared, CRA introduces additional obligations, particularly in the areas of documentation, vulnerability notification, and timely update readiness, that must be addressed now to ensure compliance by December 2027. The findings and recommendations presented in this paper offer practical guidance to help industrial equipment manufacturers and other organizations facing similar challenges prepare effectively for CRA compliance.

\section{Acknowledgements}
The authors gratefully acknowledge the partial funding provided by the Business Finland Veturi project \textit{CybersecuriTy Assurance for IEC62443 Based Environments (CTAC)}. We extend our sincere thanks to all the members of the CTAC project for their invaluable contributions. Additionally, we would like to express our gratitude to the participants of the study for their insightful feedback on CRA and the challenges it presents.

%
%
%
\bibliographystyle{splncs04}
\bibliography{citations}
\end{document}